\documentclass[a4paper,prb,twocolumn]{revtex4}
\usepackage{amsfonts}
\usepackage{graphicx}
\usepackage{amsmath}
\usepackage{amssymb}
\usepackage{latexsym}
\usepackage{psfrag}

\begin{document}

\title{Band-gap engineering and ballistic transport in corrugated graphene
nanoribbons}
\author{S. Ihnatsenka$^{1}$, I. V. Zozoulenko$^{2}$, and G. Kirczenow$^{1}$}
\affiliation{$^{1}$Department of Physics, Simon Fraser University, Burnaby, British
Columbia, Canada, V5A 1S6\\
$^{2}$Solid State Electronics, ITN, Link\"{o}ping University, 601 74, Norrk%
\"{o}ping, Sweden}

\begin{abstract}
We calculate the band structure and the conductance of periodic corrugated
graphene nanoribbons within the framework of the tight-binding $p$-orbital
model. We consider corrugated structures based on host ribbons with armchair and zigzag edges
and three different types of corrugations (armchair edges, zigzag edges as
well as a rectangular corrugation). We demonstrate that for armchair
host ribbons, depending on the type of corrugation, a band gap or low-velocity minibands appear near the charge neutrality point. For higher
energies the allowed Bloch state bands become separated by mini-stopbands. By contrast,
for corrugated ribbons with the zigzag host, the corrugations
introduce neither band gaps nor stopbands (except for the case of the
rectangular corrugations). The conductances of finite corrugated ribbons are
analyzed on the basis of the corresponding band structures. For a
sufficiently large number of corrugations the conductance follows the number
of the corresponding propagating Bloch states and shows pronounced
oscillations due to the Fabry-Perot interference within the corrugated
segments. Finally we demonstrate that edge disorder strongly affects the
conductances of corrugated ribbons. Our results indicate that observation
of miniband formation in corrugated ribbons would require clean,
edge-disorder free samples, especially for the case of the armchair host
lattice.
\end{abstract}

\maketitle


\section{Introduction}

One of the most fascinating recent discoveries in condensed matter
physics is the fabrication and demonstration of electrical conductance in
graphene, - a single sheet of carbon atoms arranged in a honeycomb lattice.
\cite{Novoselov} This discovery has ignited a tremendous interest of the
research community not only because of the new exciting physics that the
graphene exhibits, but also because of the promise of graphene-based
high-speed electronics.\cite{Adam}

While earlier works on graphene have been focused on bulk samples, 
recent studies have also addressed material, electronic and transport
properties of \textit{confined} structures such as nanoribons,\cite%
{Chen,Han07,Li} nanoconstrictions,\cite{Molitor} quantum dots\cite%
{Miao,Russo,Stampfer} and antidots.\cite{Shen} The confinement and the
patterning of graphene with few nanometer precision is typically
achieved by means of electron-beam lithography and etching techniques. 
Particular attention has been paid to control of the morphology, geometry
and stability of the device edges.\cite{Jia,Girit} These advances in 
graphene material technology and device processing make it possible to
fabricate ballistic periodic graphene structures with nanometer feature
sizes.

Periodic quasi-one-dimensional (quasi-1D) systems defined in conventional
two-dimensional electron gas (2DEG) heterostructures such as corrugated
channels\cite{Kouwenhoven,Park,Lent}, ballistic one-dimensional
supperlattices\cite{Ulloa} and arrays of quantum dots\cite{Shao} and antidots%
\cite{Ensslin,antidot} have been the subject of intense research during the past
twenty years. These artificial finite crystal structures show a wealth of
phenomena related to the formation of miniband structure and quantum
interference. Recently, a periodic structure defined in graphene, - an
antidot lattice, - has been studied theoretically by Pedersen \textit{et al}%
.\cite{Pedersen} They demonstrated that the antidot lattice can turn 
semimetalic graphene into a gapped semiconductor, where the size of the gap
can be tuned via the geometry of the lattice. An experimental realization of
the antidot array has been recently reported by Shen \textit{et al}. who
observed the commensurability oscillations and Aharonov-Bohm
oscillations arising from the artificially imposed lateral potential
modulation.\cite{Shen}

Motivated by these advances in devices processing and fabrication, in the
present paper we address the electronic and transport properties of periodic
corrugated ballistic nanoribbons. The ballistic nanoribbons represent the
fundamental building blocks of graphene-based nanocircuits and/or individual
devices. Understanding the factors that affect the electronic and transport
properties of graphene nanoribbons and exploring ways to control these
properties through periodic corrugations are the central aims of our
study.

The paper is organized as follows. In Sec. II we present a model of
corrugated graphene nanoribbons and briefly outline the basics of our
calculations of the band structure, Bloch states and the conductance based
on the recursive Green's function technique. In Sec. III we calculate the
band structures of zigzag and armchair nanoribbons with different
corrugations and discuss formation/suppression of the band gap as well as the
formation of stopbands. We calculate the conductances of finite
corrugated nanoribbons and discuss them on the basis of the corresponding band
structures. Finally, we address the effect of the edge disorder on the
conductances of realistic corrugated nanoribbons. The main conclusions of our
study are presented in Sec. IV.

\begin{figure}[tbh]
\includegraphics[keepaspectratio,width=\columnwidth]{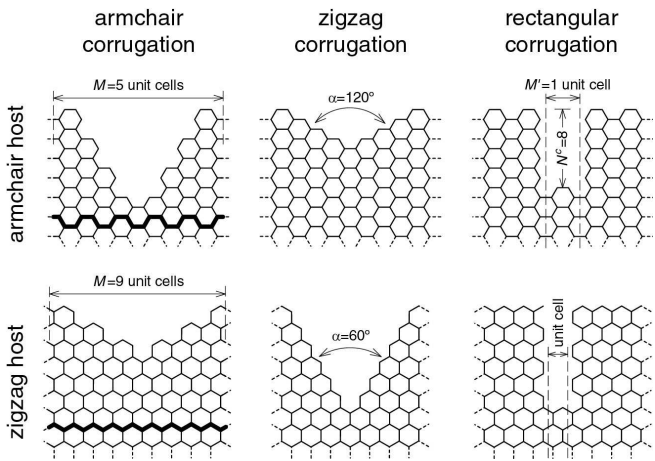}
\caption{Schematic representation of corrugations studied in the present paper. $M$ is the corrugation unit cell. $M^{\prime}$ and $N^c$ are the length and depth of a rectangular groove. }
\label{fig:diagr}
\end{figure}

\section{Model}

\begin{figure*}[th]
\includegraphics[keepaspectratio,width=\textwidth]{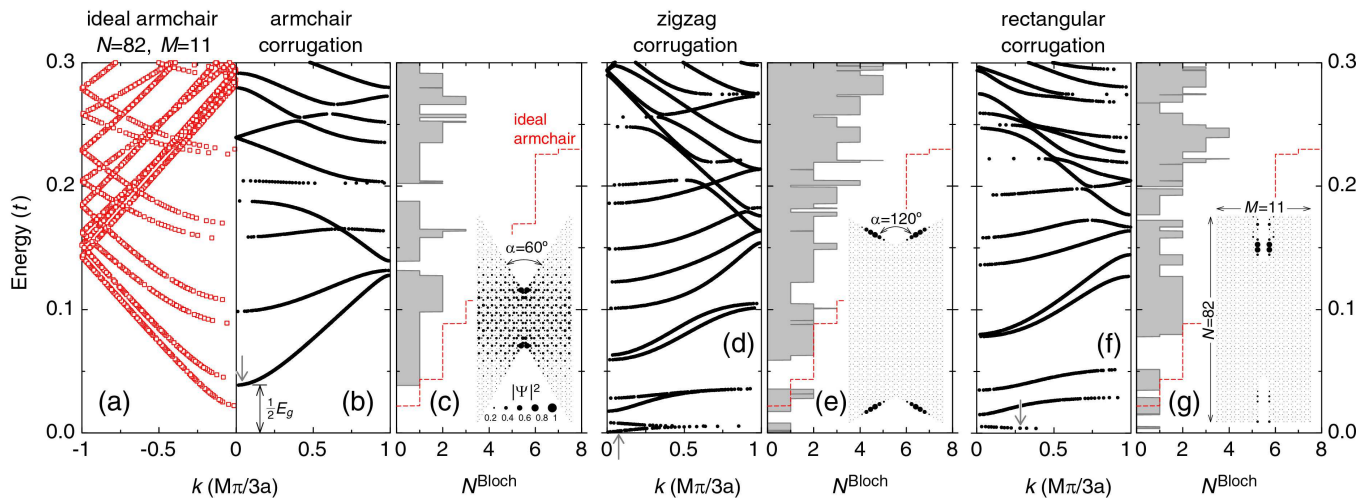}
\caption{(color online) Band structure of ideal (a) and corrugated
(b),(d),(g) graphene ribbons with the armchair host. The band gap $E_g$ is
increased if a corrugation with armchair edges is imposed (b). However
low-energy mini-bands appear if zigzag edges are present in the corrugation
(d),(f). They are associated with electron localization along the zigzag
edges as shown in the insets in (e) and (g); the wave function modulus $%
\left|\Psi\right|^2$ is given for $k$-vectors marked by arrows. The plots
(c),(e),(g) show the number of the propagating Bloch states $N^{Bloch}$: the
dashed line corresponds to the ideal ribbon, while the gray solid curve with
the filled area presents the corrugated case. The host ribbon has $N=82$
carbon atoms in the transverse direction ($\sim10$ nm width) and $M=11$ unit
cells in the longitudinal direction. $t=2.7$ eV. $a=0.142$ nm.}
\label{fig:armch_band}
\end{figure*}

We describe a graphene ribbon by the standard tight-binding Hamiltonian on a
honeycomb lattice,
\begin{equation}
H=\sum_{i}\epsilon _{i}a_{i}^{\dag }a_{i}-\sum_{\left\langle
i,j\right\rangle }t_{ij}\left( a_{i}^{\dag }a_{j}+h.c.\right) ,  \label{eq:H}
\end{equation}%
where $\epsilon _{i}$ is the on-site energy, $\epsilon _{i}=0$ in the
following, and $t_{ij}=t=2.7$ eV is the overlap integral between
nearest-neighbor atoms. This Hamiltonian is appropriate for graphene with
one dangling $p_{z}$ orbital per carbon atom and is known to describe the $\pi$ band dispersion well at low energies.\cite{Reich02} Spin and electron interaction effects are outside of the scope of our study. The effects
of corrugations are incorporated by removing carbon atoms and setting
appropriate hopping elements $t_{ij}$ to zero. It is assumed that atoms at
the edges are attached to two other carbon atoms and passivated by a neutral
chemical ligand, such as hydrogen. There are two especially simple classes
of corrugations: those with armchair edges and those with zigzag
edges. Depending on the edge type and orientation of the host ribbon, an apex
angle will equal $\alpha =60^{\circ }$ or $120^{\circ }$, see Fig. \ref{fig:diagr} and the insets
in Figs. \ref{fig:armch_band} and \ref{fig:zz_band}. For a given
edge, host and $\alpha $, it is possible to change the corrugation size by
varying its length $M$. As $M$ becomes
longer the groove penetrates deeper into the host material until only a
single carbon-carbon link persists in the constriction. As a motivation
for this choice of corrugation we refer to the recent studies\cite{Jia,Girit}
demonstrating controlled edge reconstruction with the formation of the sharp
stabilized zigzag or armchair edges. For the sake of completeness,
we consider also a third type of the corrugation, which has a rectangular
shape. Its distinctive feature is presence of the zigzag (armchair) edges
joining the armchair (zigzag) edges of the wide and narrow regions.

The Bloch states of the corrugated ribbons are calculated using the Green's
function technique.\cite{Igor08} We consider infinitely long graphene
ribbons with imposed periodic corrugation in the longitudinal direction. For
a translationally invariant system, the solutions of the Schrodinger
equation with the Hamiltonian \eqref{eq:H} obey the Bloch theorem:
\begin{equation}
\Psi_{m+M} = e^{ikM} \Psi_m,  \label{eq:Bloch}
\end{equation}
where $k$ is the Bloch wave vector and $\Psi_m$ is the Bloch wave function
at coordinate $m$. Introducing the Green's function $G=(E-H)^{-1}$ and using %
\eqref{eq:Bloch}, we formulated the eigenproblem as stated in Ref. %
\onlinecite{Igor08} and then solved it numerically. Knowledge of the Bloch
states allows one to construct the band diagram, which in turn serves as a
basis for analysis of transport properties of the periodically corrugated
graphene ribbons.

\begin{figure}[th]
\includegraphics[keepaspectratio,width=\columnwidth]{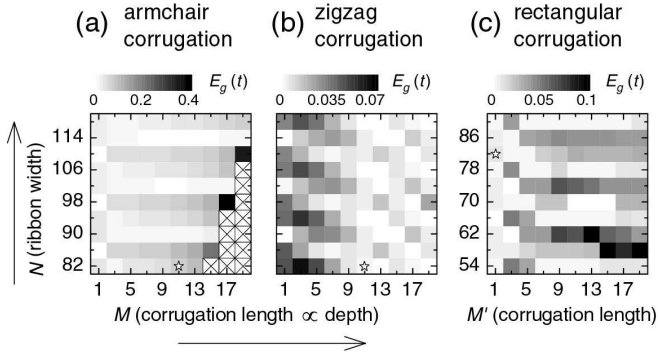}
\caption{The band gap $E_g$ vs. width $N$ of the host ribbon and length $M$ of the corrugation for the armchair host. Increase of $M$ leads to progressive corrugation
deepening and constriction narrowing for both the armchair and zigzag edges
of corrugation. Diagonal crosses indicate where the constrictions have been completely closed off. In the case of the rectangular corrugation, its depth is
fixed to be $N^c=18$ carbon atoms while the length $M^{\prime}$ increases (note
that $M^{\prime}$ is the slit width and the length of the unit cell of the corrugation equals $M^{\prime}$ plus 10 unit cells of the host armchair ribbon, see inset in Fig. \ref{fig:armch_band}(g)). Stars mark $N=82$ and $M=11$ setups shown in Fig.
\protect\ref{fig:armch_band}.}
\label{fig:E_g_vs_N_M}
\end{figure}

\begin{figure*}[t]
\includegraphics[keepaspectratio,width=\textwidth]{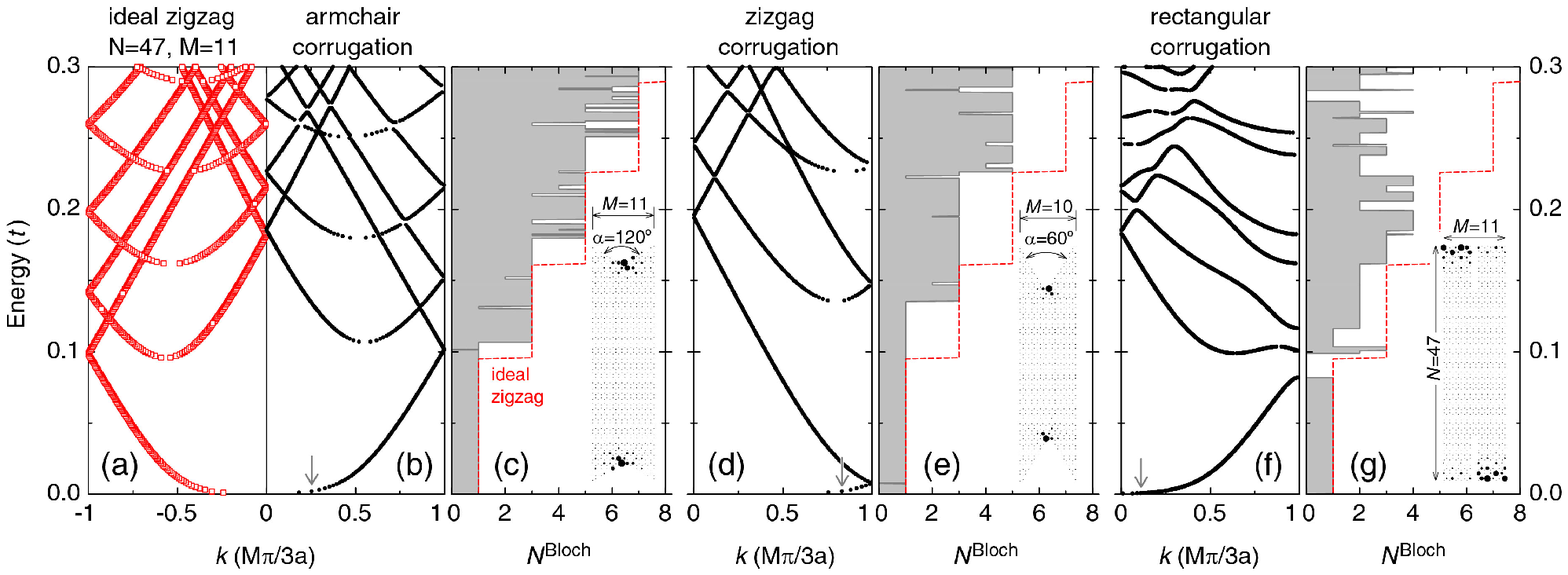}
\caption{(color online) The same as Fig. \protect\ref{fig:armch_band} but
for the zigzag host ribbon. The host ribbon has $N=47$ carbon atoms in the
transverse direction ($\sim10$ nm width) and $M=11$ unit cells in the
longitudinal direction except (d) and (e) where $M=10$.}
\label{fig:zz_band}
\end{figure*}

To analyze the transport properties, we consider $n$ periodically repeated
corrugations attached to ideal semiinfinite leads. The semiinfinite leads
consist of graphene ribbons with the same edge orientation as the host
orientation of the corrugated region. As the number of corrugations
increases, $n>>1$, we may expect the transport properties of the
corrugated channel to be governed by the Bloch states.\cite{Lent, antidot} In
the opposite limit of $n=1$, the transport is determined by scattering and
localization at the edges of a single constriction.\cite{Palacios06}

The central quantity in transport calculations is the conductance. In the
linear response regime for zero temperature, it is given by the Landauer
formula
\begin{equation}
G=\frac{2e^{2}}{h}T,  \label{eq:Landauer}
\end{equation}%
with $T$ being the total transmission coefficient. The transmission
coefficient is calculated by the recursive Green's function method, see Ref. %
\onlinecite{Igor08} for details. Note that knowledge of the Green's function
allows one also to obtain other useful information like wave functions,
density of states and currents.\cite{Igor08,Datta_book}

\section{Results and discussion}

Figures \ref{fig:armch_band}(b),(d),(f) show representative band structures of infinite
periodic graphene ribbons with different types of corrugations. All of them
are created from the armchair host of $N=82$ carbon atom width and $M=11$
unit cell length. These sizes correspond approximately to 10 nm width and 5
nm length. For reference purposes, figure \ref{fig:armch_band}(a)
presents the band structure of the ideal host armchair ribbon. It has a band gap
of $E_{g}=0.042t$ as expected for $N=82$.\cite{Dresselhaus96,Ezawa06} When
the corrugation is imposed, the band diagram undergoes substantial changes.
First, for the case of the corrugation with the armchair edge the band gap
substantially changes (in the case under consideration it increases
by a factor of 1.8), Fig. \ref{fig:armch_band}(b). However, for the
zigzag edge corrugation, conductive minibands appear near the charge
neutrality point $E=0$, Figs. \ref{fig:armch_band}(d),(f). These minibands
are very flat indicating their low velocity. Inspection of the wave
functions demonstrates that electrons in the low lying minibands are strongly localized to the zigzag
edges. Secondly, different bands of Bloch states become separated by gaps or
so-called mini-stopbands.\cite{Benisty09} Their nature is related to the
Bragg reflection due periodic perturbation.\cite%
{Kouwenhoven,Lent,Ulloa,Davies_book} Thirdly, the number of propagating
Bloch states decreases at most energies in comparison to the ideal ribbon, Figs. \ref%
{fig:armch_band}(c),(e),(g). Fourthly, in many cases avoided crossings of different minibands result in abrupt drops in the number of propagating states in narrow ranges of energy.

We also performed systematic calculations of the band structures of
corrugated ribbons of different width $N$ and periodicity $M$. 
While all of them exhibit similar behavior as outlined above for a
representative ribbon of $N=82$ and $M=11$, particular
features of the band structure depend in a sensitive way on the width of the
host ribbon $N$ and the corrugation periodicity $M$. 
This is illustrated in Fig. \ref{fig:E_g_vs_N_M} that shows the size of the
band gap as a function of the corrugation strength for the armchair host
ribbon. In the  interval of $N$ considered, the host ribbons of width $N=86,98,110$ are metallic; all other are semiconducting. For
the case of the armchair corrugation, $E_{g}$ gradually increases
as the constrictions narrow (i.e. $M$ increases) for most values of $N$, Fig. \ref{fig:E_g_vs_N_M}(a). However, for a sequence $N=90,102,114$ (corresponding to the semiconducting host ribbons), the band gap
slowly decreases and remains rather small (this periodic dependence of the $E_{g}$ on the ribbon width is clearly seen as horizontal trenches in Fig. \ref{fig:E_g_vs_N_M}(a)). This can be explained by the fact that for
this sequence the width of the corrugated ribbon in its narrowest part, $N_{narr}$, shows metallic behavior for all $M$ (i.e. the corresponding uniform ribbon of constant width $N_{narr}$ is metallic\cite{Dresselhaus96,Ezawa06}; note that $M$ is odd in Fig. \ref{fig:E_g_vs_N_M}(a)), while for all other $N$, the minimal width $N_{narr}$ corresponds to the semiconducting behavior. For the case of zigzag corrugation, the dependence of $E_g$ on the corrugation strength $M$ exhibits the opposite trend, Fig. \ref{fig:E_g_vs_N_M}(b). We attribute this to the nature of zigzag edges, which favor electron propagation at low energy. Thus for zigzag corrugations the gap tends to be larger for small $M$, for which the zigzag corrugations are interrupted most often by armchair ''defects`` that occur at each apex of the corrugated structure for zigzag corrugation of the armchair host as can be seen in Fig. \ref{fig:diagr}. The
rectangular corrugation has many similarities with the armchair case, though
electron interference at sharp corners becomes much more important. If the
constriction is long enough, substantial band gaps might develop.

Let us now consider the case of the zigzag orientation of the host ribbon, Fig. %
\ref{fig:zz_band}. The zigzag edges of graphene ribbons are known to
accommodate exponentially localized edge states.\cite{Dresselhaus96}
Independent of the width, zigzag ribbons are metallic with a zero band gap.
None of the corrugation types changes this feature and the ribbon stays
metallic unless the constrictions are a few atoms wide. As the corrugation deepens
into host material, the locations where the electron wave functions are localized shift accordingly,
see insets in Figs. \ref{fig:zz_band}(c),(e). Robust metallic behavior is
also related to a peculiar current density distribution, which is mainly
concentrated along the center of the ribbon.\cite{Palacios06, Nikolic07} Substantial mini-stopbands were found to de-
velop for the rectangular corrugation only, Figs. 4(f),(g).
(Though there are extremely narrow stopbands in Figs.
4(b)-(e), they are two orders of magnitude narrower and
thus unlikely to be observed experimentally.) Because of the abrupt change of the edge
configurations electrons stay localized near the edges of the wide regions in this case,
see inset in Fig. \ref{fig:zz_band}(g).

\begin{figure}[th]
\includegraphics[keepaspectratio,width=\columnwidth]{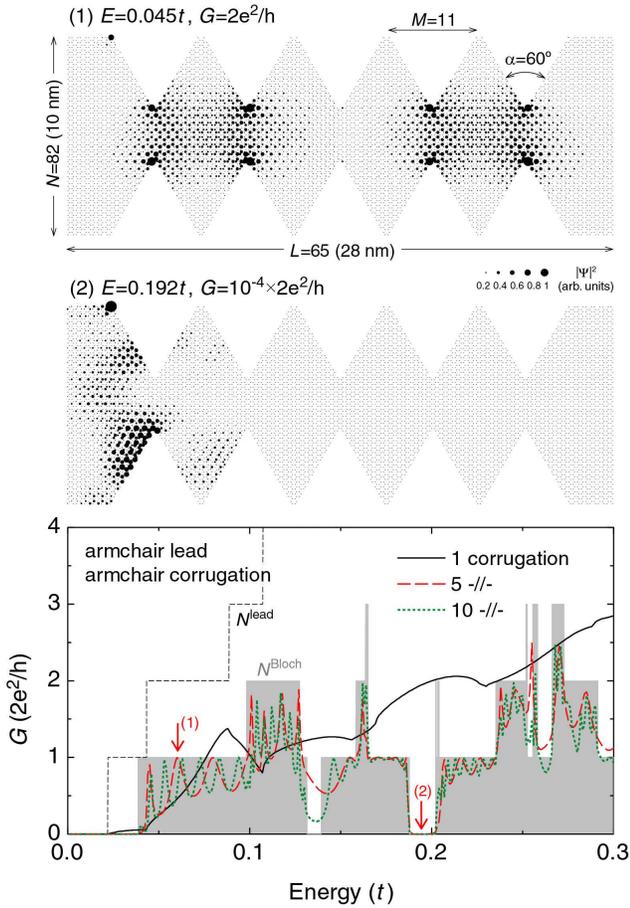}
\caption{(color online) The conductance vs. energy for the armchair ribbon with different
number of the armchair corrugations. The shaded gray area denotes the number of
propagating Bloch states for the periodic corrugation, $N^{Bloch}$, the same as in Fig.
\protect\ref{fig:armch_band}(c). The thin black dashed line shows the number of propagating
states in the leads, $N^{lead}$. The top insets present the wave function modulus for two
representative energies marked by arrows for the structure with 5 corrugations. } \label{fig:cond_armch_N_corr}
\end{figure}

\begin{figure}[th]
\includegraphics[keepaspectratio,width=\columnwidth]{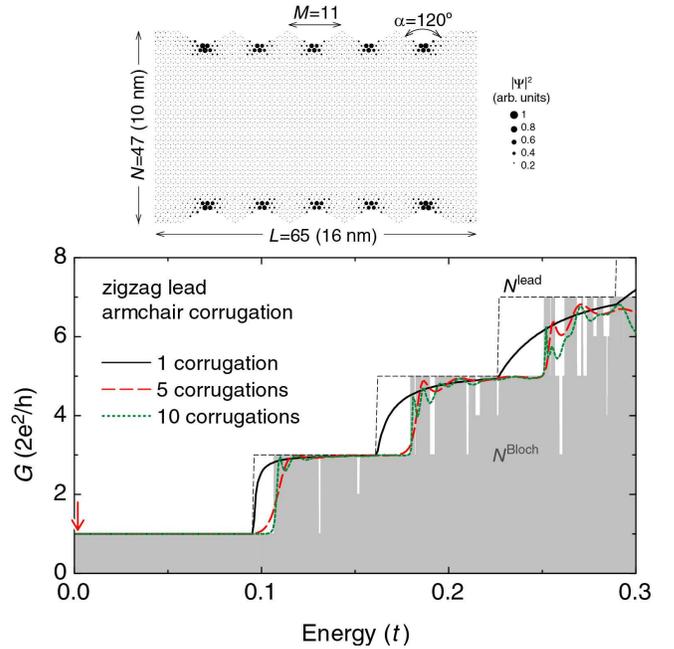}
\caption{(color online) The same as Fig. \protect\ref{fig:cond_armch_N_corr}
but for the ribbon with zigzag leads (host).}
\label{fig:cond_zz_N_corr}
\end{figure}

Let us now turn to the transport properties of the corrugated ribbons and the
relation to the Bloch states. Figures \ref{fig:cond_armch_N_corr} and \ref{fig:cond_zz_N_corr} show the conductance as a function of the Fermi energy
for different number of the corrugations. We build up the central scattering
region by connecting the corrugations in series and adding semiinfinite
ribbons, that play the role of ideal leads, to both ends. The latter has the same edge orientation as the host of
the corrugated part, e.g. they are armchair in Fig. \ref{fig:cond_armch_N_corr} and zigzag in Fig. \ref{fig:cond_zz_N_corr}. The most striking feature of these figures is the reduction of the conductance relative to that for the corresponding infinite corrugated ribbons which is given by $\frac{2e^2}{h}N^{Bloch}$, where $N^{Bloch}$ is the number of Bloch states at the Fermi energy in the infinite corrugated structure. As the number of corrugations grows the conductance follows $\frac{2e^2}{h}N^{Bloch}$ more closely. The strong oscillations observed in Fig.
\ref{fig:cond_armch_N_corr} are caused by the Fabry-Perot interference of electron waves
inside the scattering region, see the topmost inset for the wave function modulus. If the
energy of the incident electron falls into the stop-miniband the conductance may be suppressed by many
orders of magnitude  due to destructive interference as is seen at energies near $0.2t$ in the lower panel of Fig. \ref{fig:cond_armch_N_corr}. These features in the conductance of the
corrugated ribbons are very similar to those of quasi-1D periodic systems defined in
conventional 2DEG heterostructures.\cite{Kouwenhoven, Lent, Ulloa, antidot}. As is seen in Fig. \ref{fig:cond_zz_N_corr}, the electron
states existing in the zizgag ribbons have peculiar properties making them relatively immune to
the corrugations: the conductance deviates very little from $\frac{2e^2}{h}N^{Bloch}$. The wave functions effectively round the grooves and enhanced
localization of $\left|\Psi\right|^2$ becomes pronounced at the apexes, Fig. \ref{fig:cond_zz_N_corr}. This behavior resembles somewhat that of the edge states in the quantum Hall regime.

Note that our single-particle conductance calculations do not
account for charging (Coulomb blockade) effects. However, inspection of
the wave function in the corrugated ribbons for the zigzag host near the
charge neutrality point $E=0$ (and the corresponding Bloch states)
indicates that charging effects may dominate the conductance in this
regime. Indeed, because of the strong electron localization near its apexes,
the corrugated ribbon effectively behaves as an array of weakly coupled
quantum dots, see Figs. \ref{fig:zz_band}, \ref{fig:cond_zz_N_corr}. This
behavior is also reflected in a very low group velocity of the corresponding
energy bands. Thus, the corrugated ribbon with the zigzag host is expected
to effectively function as an array of weakly coupled quantum dots, even
though the corrugations can be relatively small in comparison to the
ribbon's width.

\begin{figure}[th]
\includegraphics[keepaspectratio,width=\columnwidth]{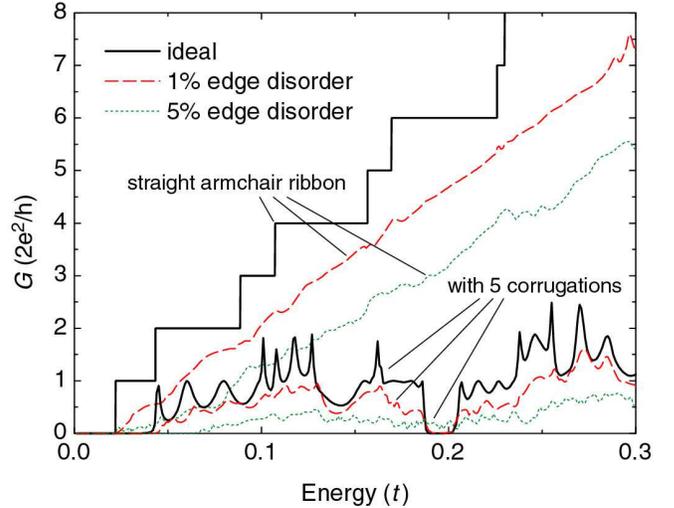}
\caption{(color online) Influence of disorder on the conductance through
both the straight armchair ribbon and the ribbon with 5 armchair corrugations. The
structures have the same width $N=82$ ($\sim10$ nm) and length $L=65$ ($%
\sim28$ nm). }
\label{fig:disorder}
\end{figure}

It is well known that the disorder inevitably present in realistic devices
has a marked effect on the conductance.\cite{Han07, Martin08} Figure \ref%
{fig:disorder} shows representative results for both a corrugated ribbon
and the straight ribbon of the same size. The disorder is introduced onto
the edges only and the host ribbon and corrugations are both armchair. We
averaged the conductance over ten realizations of the disorder in order to
reduce fluctuations and facilitate visualization of the results. The edge
defects act as randomly positioned short-range scatterers and induce strong
backscattering. It leads to Anderson localization with substantial transport
gaps opened for narrow ribbons.\cite{Han07, Martin08, Lewenkopf} Even 1\%
of defects is enough to destroy conductance quantization for the straight
ribbon and the Fabry-Perot oscillations for the case of the corrugated
ribbon, Fig. \ref{fig:disorder}. The stopbands are smeared already
for 5\% edge disorder, which indicates that the observation of the
miniband formation in the armchair-host corrugated ribbons would require
extra-clean, edge-disorder free samples. Note that we also performed
calculations for disordered ribbons with the zigzag host. As expected, they
showed behavior similar to that shown in Fig. \ref{fig:disorder} though the defect concentrations producing similar effects on the conductance are
significantly larger.

\section{Conclusions}

We studied the electronic and transport properties of periodic corrugated graphene
nanoribbons within the standard $p$-orbital tight-binding model. We considered both 
armchair and zigzag underlying host ribbons and three different types of corrugation
defined by grooves with armchair edges, zigzag edges as well as a rectangular
corrugation. We calculated the dispersion relations and Bloch states for corrugated
ribbons, and demonstrated that they exhibit different features depending on topology of
the host ribbon (i.e. armchair or zigzag). For the armchair host, depending on the
type of corrugation, a band gap or low-velocity minibands appear near the charge
neutrality point $E=0$. For higher energies bands of Bloch states become separated by
mini-stopbands. By contrast, for corrugated ribbons with the zigzag host, the
corrugations introduce neither a band gap nor substantial stopbands
(except for the case of the rectangular corrugations).

We calculated the conductance of graphene ribbons with finite numbers of
corrugations $n$. As expected, for sufficiently large $n$ the conductance
follows the number of the corresponding propagating Bloch states and shows
pronounced oscillations due to the Fabry-Perot interference within the
corrugated region. We also argue that for low electron energies the corrugated ribbon with the
zigzag host is expected to effectively function as arrays of weakly
coupled quantum dots with the conductance dominated by the single-electron
charging effects, even though the corrugations can be relatively small in
comparison to the ribbon's width.

We also demonstrated that as in the case of uncorrugated ribbons, edge
disorder strongly affects the conductance of the corrugated ribbons. Our
results indicate that observation of miniband formation in corrugated ribbons would require extra-clean edge-disorder free samples,
especially for the case of the armchair host lattice.

Finally, we hope that our study will motivate further experimental
investigation of periodically corrugated graphene nanoribbons.

\begin{acknowledgments}
This work was supported by NSERC and The Canadian Institute for Advanced Research. I.V.Z. acknowledge financial support from the Swedish Research Council (VR).
\end{acknowledgments}


\end{document}